\documentstyle[12pt]{article}
\begin{document}
\title{On the selection of preferred consistent sets} 
\author {C. Anastopoulos \\ Theoretical Physics Group, The Blackett Lab. \\
Imperial College \\ E-mail : can@tp.ph.ic.ac.uk 
Imperial TP/96-97/65}
\date { September 1997}\maketitle

\begin{abstract}
The theme of this paper is  the multiplicity of the consistent sets emerging 
in the consistent histories
approac to quantum mechanics. 
We propose one criterion for
 choosing preferred families among them: that the 
physically realizable quasiclassical domain ought to be one corresponding 
to classical histories. We examine  the way classical mechanics
arises as a particular window, and in the important role played by the canonical 
group.  We finally discuss a possible implication of our selection criterion:
that only a class of Hilbert space operators correspond to physical quantities 
and hence the full non-distributivity of the lattice of quantum propositions 
is redundant.
\end{abstract} 
\par
The consistent histories approach \cite{Gri,Omn1,Omn2,GeHa, Har, Ish12, DH,DK}
(or rather approaches taking into 
account its diverse existing formulations ) has been a subject of 
increasing interest in recent years. It provides a view of time as 
an intrinsic object in the quantum theory (particularly appealing when 
considering quantum cosmology), a natural language for  discussion of the 
emergence of classical behaviour and is believed to constitute a converging
 point for diverse ideas from different interpretations of quantum theory.
\par
Still it seems that an old problem (or at least a point of unease) of the 
 Copenhagen interpretation has been transferred into the histories approach:
 the notion of complementarity. The inability according to the Copenhagen 
interpretation to assign simultaneously properties to a quantum system that 
 correspond to non-commuting operators, is translated in the history language
 as the question of the many different ``windows to reality'' or more precisely
 the multitude of maximal consistent sets of histories corresponding to one 
decoherence functional. Both are an expression of the non-distributive 
character of the lattice of quantum propositions; or equivalently from the
 implicit fundamental assumption that all physical observables (and consequently
 propositions about them) have a counterpart in Hilbert space operators.
\par
The counter-intuitive ( and rather disturbing ) nature of  the attitude to accept all 
 windows as real has been demonstrated in a series of papers by Kent \cite{Ken1,Ken2}.
 He showed, in particular  that a) contradictory propositions can be inferred
 with probability one when one is reasoning within different consistent sets and
 b) a quasiclassical domain is generically either not stable in time or 
 does not allow inferences according to the predictions of the corresponding classical theory.
 In absence of a precise selection criterion, the consistent histories theory loses 
much of its predictive power, since for instance the quasiclassical limit of the 
non-relativistic classical mechanics cannot be uniquely determined. We thus lose 
sight of the transition `` from Hilbert space to common sense''.
\par 
In this paper we  will argue under the assumption  that somehow only
 one window is physically relevant and realizable 
, the one corresponding to the 
classical world as we experience it.
\footnote{ This is not the only option , see for instance \cite{topos}}
 There has been a number of ideas and conjectures
 for an algorithm that would enable us to select this particular window 
\cite{Kentsel,IL}, while
 keeping the established mathematical structures of the history formalism. 
I think that we should distinguish two possibilities in the formulation of
 such a criterion: \\
i)   introduce an additional postulate (intrinsic in the mathematical structure 
of the theory)\\
ii) specify the physical context of a history theory and identify within this a
 mechanism that selects a physical window.\\
We should elaborate on this point in order to clarify the distinction.
 History theory in its current axiomatic form is of  a purely logical - statistical  nature. 
Its primitive concept is the notion of temprorally ordered propositions, 
it contains a rule for probability assignement and implication is defined through
 conditional probability. It is a universal theory since in no point in its axiomatics
 does the need  to specify its physical content arise. In this sense history theory
 (as any quantum logic scheme) can be viewed as the attempt to capture the general rules,
 upon which the reasoning about physical systems should be based. And of course,
 this raises the question (implicit in many debates about the completeness of quantum 
mechanics) whether it is meaningful to describe the physical world using 
 a context-free abstract language. This debate is particularly pronounced here, since 
the many windows issue might be taken to imply that not even a 
concrete representation of the logic can be sufficient to determine the physical 
content of a theory.
\par 
The difference between the two lines of inquiry should  now be clear. According to the
 first, there would be no reason to abandon the universal,  logical character of the
 histories axiomatics, while the second admits the necessity to seek a selection mechanism 
in the physics of the system. 
\par
 Proposals of type (i) discussed or implied so far are mainly of a``teleological''
\footnote{In the sense that minimum entropy or minimum action principles are.} type:
 minimization of some information measure or maximization of some sharp probability
 measure defined on sets of consistent histories. I do not think that a criterion of
 this type can be eventually succesful. The reason for this lies in the plethora of 
consistent sets, many of them being remarkably trivial. For instance, if one wanted
 to use minimization of the IL entropy as a criterion (which is not something the
 authors of \cite{IL} have proposed but could be  a valid conjecture) we would have to
 contend with histories made out of spectral projections of the initial density matrix.
 But even if a criterion nicely avoiding all trivial cases were to be found, we would
 still have to cope with the issue of the non-persistence/ non-predictability of the
 selected quasiclassical domain. To avoid this one would have to consider sets of histories
 with temporal support reaching arbitrarily far into the future. This would entail a highly
 uncausal prediction algorithm, in direct conflict with the (cosmological) 
motivations of a history theory.
 \par 
If, on the other hand, we look for a universal physical selection mechanism,
 we should not have to think much, before conjecturing that 
the ever-present gravitational field might be somehow relevant. This would be a history
 language transcription for Penrose's idea of gravitationally induced state
 vector reduction. Appealing as it might seem, it still has its problems; if a 
quantum theory of gravity exists and is to be writen in the history language,
 what would be its preferred quasiclassical domain? (Or are we to consider quantum gravity
 as the realm of the many coexistent windows?) 
\par 
The above remarks have set the conceptual ground, upon which the issues discussed in
 this paper are to stand. More precisely our aim is as follows:  a) propose a natural minimal condition to be satisfied by 
a preferred window, b) examine a concrete case :the emergence  of classical mechanics , c) use the insight obtained by this 
 to strngthen our condition into  a sharp selection criterion. The discussion will suggest the  existence of preferred variables in quantum mechanics. Finally by inverting the argumentation 
we will be led to 
 examine the possibility of formulating a history theory with reference
  only to those objects. Such a version of history theory, would not need to keep the
 full non-distributivity of the lattice of quantum propositions and would seem to converge
 to strongly realist interpretational schemes for quantum mechanics.
\par 
 We start by stating what the minimum requirement for a selection criterion can be and through the 
study of their consequences we shall phrase we shall arrive at a more precise formulation. Our minimum requirement is that {\it there is a sense according to which the preferred window retains the temporal 
structure of the full history theory}. More precisely we want the pre3ferred windowe to correspond to 
a classical history.
\footnote { This requirement seems to be a type (i) one, for it can be formulated solely 
in terms of primitive elemetns of the theory ; its underlying assumption is that there 
is no  sense in reasoning within windows that do not respect the temporal structure of 
the full theory. But we shall see that if it is to give a predictable theory a commitment to the physical content (though rather slight) should be made}
 By window to reality
 we define a maximal subset of the algebra of history propositions, such that any 
exhaustive and exclusive set constructed out of its elements is consistent.
Now, this criterion, is an implicit one in Gell-Mann and Hartle's definition of a 
quasiclassical domain as a maximal consistent subalgebra on which approximately deterministic
 laws can be defined. Of course this approach  is in conflict with ours, since is meant 
to identify the nature of the window attuned to which an intelligent data gathering
 and computing system can evolve \cite{GeHa}, without doubting the simultaneous validity of the
 different windows (this  has 
been discussed extensively in \cite{DK})
\par
Now we proceed to state more precisely what is involved in the above requirement.
 For this, one needs first to define what we mean by classical history theory.
\\ \\
{\bf Classical histories}\\
We will prefer to cast the classical histories in the language of classical probability
 rather than classical mechanics (CM), for the logical structure of the latter is a special
 case of the former.
\par
The primitive element in probability theory is the notion of the sample space:
 a measure space $X$ (can be taken as a symplectic manifold in the case of CM).
 The space of observables is the commutative algebra $O(X)$ of  random variables
  i.e  measurable functions from $X$ to ${\bf R}$ (usually taken to satisfy some suitable condition
 according to extra structures on $X$
(e.g. smoothness in the case of CM) A state $\omega$
 is a positive linear functional on $O(X)$ assigning to each random variable $A$ its mean 
value $\omega(A)$ ( it corresponds to  a positive element $\rho_{\omega}$ of ${\cal L}^1(X)$. 
\par
A single time proposition corresponds to  a measurable subset $C$ of $X$ ( the system
 lies in $C$ at this particular time). The set of all measurable subsets of $X$ forms a
 Boolean lattice ${\it B}(X)$, which can be concretely represented by the projective elements
 of the Banach algebra ${\cal
L}^{\infty}(X)$: these are nothing but the characteristic 
functions $\chi_C(x)$  of the measurable sets. The association 
of the characteristic functions 
to propossitions about values of random variables is established through the commutative 
algebra version of the spectral theorem
\begin{equation}
F(x) = \int \lambda d \lambda \chi_{C_{\lambda}}(x)
\end{equation}
where $F(x)$ a general element of $O(X)$ and $C_{\lambda}$ is the subset of $X$ where 
$F$ takes value $\lambda$.
\par 
It is easy to see how one can translate the constructions of quantum mechanical histories
 in this case. An $n$-time history is essentially a string of measurable subsets of
 $x$: $(C_{t_1},C_{t_2},\ldots,C_{t_n})$. This is represented by the projector 
$\chi_{C_1} \otimes \dots \otimes \chi_{C_n}$, which is an element of
 $\otimes_n {\cal L}^{\infty}(X) = {\cal L}(\times_n X)$. The construction of the space
 of history propositions then trivially proceeds along the lines of Isham \cite{Ish12}. 
\footnote{ The construction for continuous-time histories should follow similar reasoning:
 taking the quantities $x_t \in \times_{t}X_t$ as corresponding to fine grained histories,
 and constructing coarse-grainings through the use of a Wiener measure on $X$.}
\par 
In a classical theory there is no obstacle in consistently assigning probability measures to 
all histories. Nevertheless, we will try to write the probability assignment through
 a decoherence functional, to establish contact with the quantum case. Of course,
 the interference measured by this will be trivial, i.e. it will be due only to the
 non-disjointness of a pair of histories.  
\par 
Time evolution in a classical probabilistic theory  is given by a family (actually a
 semigroup) of bistochastic maps 
\footnote{A  linear, positive map $T:{\cal L}^{\infty}(X) \rightarrow {\cal L}^{\infty}(X)$
 is called bistochastic if $T(1) = 1$ and its restriction on ${\cal L}^1(X)$ is
trace preserving.}
 $T_t :{\cal L}^{\infty}(X) \rightarrow {\cal L}^{\infty}(X)$. These induce a family  of maps 
$T^{\dagger}_t$ on the space of states by $ (T^{\dagger}_t \omega)(A) = \omega(T(A))$, 
in terms of which a (real-valued) decoherence functional for a pair  of histories
 $\alpha = (C_1,\ldots,C_n)$ and $\alpha' = (C_1',\ldots, C_n')$ with same temporal support,
 reads
\begin{equation} 
d(\alpha,\alpha') = Tr \left(  \chi_{C_n}(T^{\dagger}_{t_n -t_{n-1}}     
( \ldots T^{\dagger}_{t_2-t_1} (\chi_{C_1} T^{\dagger}_{t_1}(\rho_0))\chi_{C_1'})
  \ldots )) \chi_{C_n'} \right) 
\end{equation}
  We should distinguish two cases:\\
i) Deterministic dynamics: $T$ is an automorphism of the algebra of observables
  and generically can be defined through a permutation $\tau$ on $X$ 
( a canonical transformation
 for the case of CM): $T(A)(x) = \tau^*A(x)=  A(\tau x)$. In this case it is trivial
 to verify that a history can be represented by a projector on ${\cal L}^{\infty}(X)$, 
namely
\begin{equation} 
 \chi_{\alpha} = \chi_{\tau_1 C_1 \cap \tau_2 C_2 \cap \ldots \tau_n C_n}
\end{equation}
ii) Random dynamics: $T$ is a convex combination of automorphisms: $T = \sum_i \lambda_i
 \tau_i^*$. Using the convexity property of the decoherence functionals \cite{IL},
 it is easy to
 establish that expression (3) defines a well behaved decoherence functional.
\footnote{Incidental to our aims, is the fact that this construction of a decoherence 
functional containing random dynamics 
 can be repeated (modulo a few technicalities) in the quantum mechanical case by making
 use of the ILS theorem \cite{ILS}. In particular this implies that theories of stochastic state
 vector reduction (like in \cite{GRW} ) can be nicely incorporated in the history
 formalism, and hence suffer from  the multiple windows problem. We can therefore conclude
  that no modification in the dynamics, is  sufficient by itself to provide a selection 
criterion.} 
\par
Finally we choose to define implication through conditional probability.
 If $\alpha$ and $\beta$ are disjoint propositions then we say that $\alpha$ {\it implies} 
$\beta$ if $p(\alpha \cup \beta) = p(\alpha)$. 
\par
We could then proceed and incorporate the classical history theory as a special 
case of the history axiomatics as written down by Isham 
\cite{Ish12}, except for the fact that not every Boolean lattice is suitable as
 a space of history propositions. Fundamental for any formulation of probability theory
 is the stability of the sample space through time. Hence the lattice of history
 propositions should always be identified with the characteristic functions of some
 measurable space of the form $\times_t X_t$, where the $X_t$ are related by
 structure-preserving bijections. 
\\ \\
{\bf The classical mechanics window}\\
We are now going to examine the sense in which  the world of classical mechanics arises
 as a particular window in a  quantum mechanical history theory . The aim is to
identify the important structures that eventually give the corresponding quasiclassical domain and use this information to sharpen our previous requirements to a selection criterion. We will restrict ourselves to the case of non-relativistic 
particle mechanics, since a) for quantum field theory neither do we have a history version,
 nor do we have a clear understanding of its classical limit, 
b)  we shall rely on the insight obtained by Omn\'es semiclassical theorems
 \cite{Omn2}.    
\par
Let us anticipate the discussion to follow and give a useful characterization for
 a class of classical histories. Whenever the sample space carries a metric structure one
 can define a map $\nu : {\cal B}(X) \rightarrow {\bf R}^+$, such that $\nu(C)$ gives a 
measure
of the size and regularity 
of the cell $C$   (i.e the ratio of the area of its boundary $\partial C$ to its volume), 
in such a way that $\nu(C)$ is close to zero when $C$ is sufficiently large and regular. 
There is a precise sense in which $\nu$ can be constructed; the reader is referred to Omn\'es 
\cite{Omn2} for details. We then call a classical history theory $(\times_t X_t,d,\mu)$ 
$\epsilon$-deterministic, if for any cell $C$ such that $\nu(C) = O(\epsilon)$ there exists 
another cell $C'$ with $\nu(C') = O(\epsilon)$ such that the conditional probability
$p(C,t_1;C',t_2|C,t_1)$ is of order $O(\epsilon)$. This essentially means that the 
randomness of the dynamics is significant only on scales of the order of the characteristic
length of the metric. Two $\epsilon$- deterministic theories are equivalent if they have isomorphic 
sample space $X$ and the conditional probabilities of the two theories differ in the order of $\epsilon$.
\par
Omn\'es' construction starts from the well-known fact that the classical phase space can be 
naturally embedded in the projective space of its corresponding Hilbert space by the use of 
the canonical  group. Starting from the representation of the canonical group on the Hilbert 
space of the theory, one can construct the mapping $j_r$ from the phase space $\Gamma$ 
to the coherent 
state projectors $z \rightarrow P_z = U(z)|r \rangle \langle r | U(z)^{-1}$, where $|r \rangle $ 
an arbitrary vector of $H$. Hence $j_r$ embeds $X$ into ${\cal P}H$. The pull-back 
of the Fubini-Study metric on ${\cal P}H$ defines a metric on $X$ , with respect to which 
a function $\nu$ of the type discussed above can be constructed. Note, that for a generic 
group there is an optimization algorithm \cite{Per} for the choice of $|r \rangle$, so that 
the characteristic scale of the metric on $X$ is essentially $\hbar$. In the case of quantum 
mechanics on ${\cal L}^2({\bf R}^n)$ this algorithm leads to Gaussian coherent states.
\par
One can then construct approximate projectors corresponding to "classical'' type of propositions
about phase space cells:
\begin {equation}
P_C = \int_C d \mu(z) P_z
\end{equation}
where $\mu$ is a measure on $X$. If $\nu(C) = O(\epsilon)$ , then any projector in an
 $\epsilon$-neighbourhood of $P_z$ can be thought as representing the cell $C$. 
\par
To establish consistency one needs then an important fact: that coherent states on ${\cal L}^2
({\bf R}^n)$ are approximately stable under time evolution generated by Hamiltonians of the form
$H = \frac{p^2}{2m} + V(x)$ for a large class of physically interesting potentials $V(x)$. This 
can be used to establish consistency for all exclusive and exhaustive sets of histories 
constructed out of quasiprojectors corresponding to cells $C$ with small value of $\nu$.
\footnote{ What Omn\'es has actually established is approximate consistency (of order $\epsilon = 
\nu(C)$)
of these sets, and assumed that there exist history propositions close to them, out of which 
an exact consistent set can be formed. Although this is a plausible assumption, it still remains 
to be proved. Let us see what is involved in this proof.
Taking two n-time histories the decoherence functional can be restricted to a 
continuous map 
$d: (\otimes_n B(H)) \otimes (\otimes_n B(H)) \rightarrow {\bf C}$. Consider a set of $N$ history 
propositions constructed by approximate projectors $P_i$, $i = 1 \ldots N$, and assume 
$d(P_i,P_j) < \epsilon$ ( an unsharp approximate consistency criterion ). Now if one defines the
sets $O_{\epsilon} = d^{-1}(\{z \in C; |z|< \epsilon \})$ and $C_0 = d^{-1}(\{0\})$, it is 
sufficient to show that the connected component of $C_0$ contained in the same component of 
$O_{\epsilon}$ with $P_i \otimes P_j$ has non-empty interior. For then we can use the fact that
the set of projectors is dense in $B(H)$ to associate exact projectors $P^{exact}_j$ 
in as $\epsilon$-neighbourhood of  $P_j$ such that we have exact 
consistency. This seems to a generic case for sets of $n-time$ histories in infinite dimensional 
Hilbert spaces. Of course, when one assumes continuous time histories the issue becomes more 
complicated. A study of the general case would be invaluable towards understanding 
the   "predictability  sieve'' on the possible fine-grainings 
of a consistent set.}
We also get approximate determinism, corresponding to evolution with a classical Hamiltonian 
$H(z) = Tr(P_zH)$. In this sense the window correponding to classical mechanics corresponds to
a classical $\epsilon$-deterministic history theory (by no means a unique one).
\par
From the above discussion it is clear that the classical mechanics window satisfies our previously stated requirement. We are now in a position to give a more precise characterization : {\it A preferred 
window such that each consistent subset of it can be embedded into the lattice of propositions 
of a classical history theory}. It might be the case that to get uniqueness, we must impose some 
condition of maximality on the corresponding classical history ,so that trivial windows will get excluded. 
An example is the window corresponding to propositions about values of energy, any subset of which 
is trivially consistent (it seems that one cannot get any contradictory inferences by considering 
such a set in conjunction with the CM window). Anyway, if it turns out to be necessary to include 
a maximality condition, this can 
easily take the form of specifying an extra structure (i.e. the requirement that it is 
a topological space or a manifold for the case of CM) for the sample space 
$X$ of the emergent (or underlying?) classical history theory.
\par
In our particular case, the window corresponding to classical mechanics  can be shown to 
be essentially unique. For a different selected window (corresponding to a classical history with 
sample space $X$ ), would imply the existence of an embedding of $X$ on the projective Hilbert 
space (or more generally the existence of a projection-valued measure (PVM) on $X$.
  Without loss of 
generality we can take $X$ to be a subspace of ${\bf R}^n$, and consider the marginal PVM corresponding 
to an one-dimensional submanifold of $X$. This defines a self-adjoint operator. Now a fundamental 
property of the canonical group is that its spectral projections generate the whole of $B(H)$ and
hence this operator can be represented as $f(\hat{z})$ 
, where $\hat{z}$ represents the 
generators of the canonical group and $f$ some measurable function
. This means that the classical 
history propositions corresponding to $X$ can be embedded into the lattice of  history propositions of 
the CM  window. \footnote{ There are factor ordering ambiguities when choosing $f$. To make the above arguments into a rigorous proof one needs to show that observables corresponding to fucntions with different factor ordering define equivalent $\epsilon$-deterministic thepories. This is quite plausible for
$\epsilon$ in the order of $\hbar$ in some power in view of already existent semiclassical theorems 
, but we have been unable to find a general proof}
It is in this sense that the CM window 
(assuming of course that it exists) is maximal. (Note that in the above argumentation $X$ is assumed to be a manifold).
 A remark is in order at this point concerning the role of the hydrodynamic variables 
discussed extensively by Gell-Mann and Hartle. Whenever the CM-window exists, the construction 
of theses variables proceeds along the lines of classical hydrodynamics, since they can be 
viewed as further coarse graining on the emergent classical history theory. More interesting 
would be the case where the CM-window does not exist (i.e. the propositions constructed from 
the canonical group form no non-trivial consistent sets). In that case it is conceivable that 
two complementary hydrodynamic windows might exist. This would of course invalidate our argument
that the proposed selection criterion chooses a unique window. ( It is hard to imagine  
such a set of  variables  that does not reduce to ordinary hydrodynamics or at least 
to an extension of irreversible thermodynamics). The physical system where this might be possible 
is quantum field theory, where there are indirect arguments \cite{Ana}
(though definitely not conclusive) that the classical field theory window might not be 
emerging for a large class of states of the system.
\\ \\
{\bf Preferred window and non-distributivity}
\\
From the arguments stated earlier, it should be clear that { \it if we commit ourselves to the logic of 
one preferred window}, we cannot escape the conclusion that they are generated by a particular class 
of Hilbert space operators (the ones determined by the canonical group). The question then  
arises: what is the nature of the other windows ( the ones not stable in time)? They cannot be considered
as anything but redundant, for 
they are irrelevant  to any physical predictions of the theory. But then, why do they appear at all in the
formalism? The only possible answer is that there is a redundancy in the primitive elements of the history 
theory. Not all history propositions can be considered as physical. This is a point strongly reminiscent of
the one advocated in the context 
of single-time quantum mechanics by Margenau and Park \cite{Mar}: the incomeasurability of observables corresponding 
to non-commuting operators  can be true in quantum mechanics 
{\it only if accept as an axiom that any operator 
on the Hilbert space corresponds to some physical observable.} The many windows problem is just the history counterpart 
of the incomeasurability problem of single-time quantum mechanics; and it is due to the richest structure of the former 
that the counterintuitive effects of such a postulate are most impressive.
\footnote{ Actually one of the strongest arguments posed by Margenau and Park, was the time-of-flight type of experiments, 
with an analysis that can be thought as an ancestor of the history formalism}
\par
A consequence of not accepting all operators as physical, is that there is no need to keep the non-distibutive 
character of the lattice of history propositions (see \cite{Mar} for extensive discussion); indeed reasoning inversely, the 
non-distibutivity is the main assumption that allows multiple windows. One should then start thinking of possible substitutes.
If one  opts for a distributive lattice structure for history propositions, the first choice would be to consider Boolean 
lattices (this is  not the only possible choice, see for instance \cite{Aus}). Of course, hidden variables theories fall in this category (even though they would correspond to a classical history theory). Another approach would be Sorkin's quantum measure theory, where the lattice of history propositions is assumed Boolean and the role of the decoherence functional is played bya non-additive measure on this lattice \cite{Sor}.
\par
It turns out that a history theory with a preferred window looks similar to a Sorkin-type construction. Recall the 
importance of  the canonical group towards identifying the preferred window. In standard quantum mechanics, its role is also 
important. Since all infinite dimensional Hilbert spaces are isomorphic,
 the only way one can separate the physical content of
(say) ${\cal L}^2({\bf R})$ from ${\cal L}^2({\bf R}^3)$ is by 
 considering the representations of different canonical groups ( its 
history version is also very important towards constructing explicitly the space of history propositions). One is then tempted to the question? Can one construct a history theory using only ``classical'' primitive elements (canonical group or the 
phase space that can be constructed from it, possibly plus some additional (complex) structure)? One can for instance assume as primitive elements the paths $z(t),z^*(t)$ on some ``phase space'' $X$ and their coarse grainings as given by integration with respect to some Wiener measure and a coherent-state path-integral version of the decoherence functional between pairs of coarse grained 
histories $\alpha = \int_{C_{\alpha}} d \mu(z(.),z^*(.))$ and $\alpha' = \int_{{C_\alpha'}} d \mu(z(.),z^*(.))$
\begin{eqnarray}
 d(\alpha,\alpha') = \int dz_f dz^*_f dz_i dz^*_i dz^*_i dz'_i dz'^*_i e^{-z^*_f z_f - z_i^* z_i - z'^*_i z'_i} \nonumber \\
\left( \int_{C_{\alpha}} d \mu(z(.),z^*(.)) \int_{C_{\alpha'}} d \mu(z'(.),z'^*(.)) e^{iS[z(.),z^*(.)] - i S[z'(.),z'^*(.)]}\right) \rho(z^*_i,z'_i)
\end{eqnarray}
with path integration over paths such that $ z(0) = z_i$, $z^*(t) = z^*_f$, $z'(t) = z_f$, $z'^*(0) = z'^*_i$.
Modulo some difference in the probability assignment this construction is a close relative of Sorkin's theory.
\par
Unfortunately, this construction is at least incomplete and for a more fundamental reason than the proverbial non-definability of the 
path integral with Wiener measures:  there is no way we can reproduce the quantum mechanical combination of subsystems via the tensor product solely from the knowledge 
of $X$; we have to introduce a linear structure and hence the Hilbert space would eventually enter again our schemes.
\par
Before concluding let us summarize the thesis of this paper. I one wishes for a history theory allowing a) maximum predictability and b) a 
realist commitment, the option of seeking a selection algorithm among different windows is a natural one. Our proposed selection criterion
(of temporal stability of the preferred window) seems to imply that only a class of history propositions is physically relevant, hence that the 
non-distributivity of the lattice of history propositions is redundant. If one then attempts to write an ``economic'' version of history theory, 
where only physical quantities are taken into account, then one faces one fundamental problem: how would the Hilbert space structure emerge in 
such a theory. The problem can be neatly summarized as the inverse of the many windows problem: How does one recover the Hilbert space from 
(distributive) ``common sense''?

\section{Aknowlegements}
I would like to thank A. Kent for an important discussion on this subject and encouraging me to proceed on that. Also K. Savvidou for insisting that I read reference \cite{Mar}.


\begin{thebibliography}{}

\bibitem {Gri} R.\ B.\ Griffiths,  J. Stat. Phys. {\bf 36}, 219 (1984).

\bibitem {Omn1} R. Omn\`es,  J. Stat. Phys. {\bf 53}, 893 (1988); 
J. Stat. Phys. {\bf 53}, 957 (1988);  Rev. Mod. Phys. {\bf 64}, 339 (1992).

\bibitem {Omn2} R.\ Omn\`es,  J. Stat. Phys. {\bf 57}, 357 (1989);  R. Omn\`es, {\sl The Interpretation of Quantum
Mechanics} 
( Princeton University Press, Princeton,
1996.)

\bibitem {GeHa} M.\ Gell-Mann and J.\ B.\ Hartle, in {\sl Complexity, Entropy 
and the Physics of Information}, edited by W.\ Zurek, ({\sl Addison
Wesley,Reading} 
(1990));  Phys. Rev. {\bf D47}, 3345 (1993).

\bibitem{Har}  J. B. Hartle, Spacetime quantum mechanics and the quantum mechanics of spacetime, in {\sl Proceedings of the 1992 Les Houches School, Gravitation and Quantizations}, 1993.

\bibitem{Ish12} C. J. Isham, Jour. Math. Phys {\bf 35}, 2157 (1994); C.J. Isham and N. Linden, J. Math. Phys. 5472,(1994).

\bibitem{DH} H.F.Dowker and J. J. Halliwell, Phys. Rev. {\bf D46}, 1580
(1992).

\bibitem{DK} H. F. Dowker and A. Kent, J. Stat. Phys. {\bf 82}, 1575 (1996).

\bibitem{Ken1} A. Kent, Phys. Rev. Let. {\bf 78}, 2874 (1997).

\bibitem{Ken2} A. Kent, Phys. Rev. {\bf A54}, 4670 (1996).

\bibitem{topos} C. J. Isham, Int. J. Theor. Phys. {\bf 36}, 785 (1997).

\bibitem{Kentsel} A. Kent and J. Mc Elwaine, Phys. Rev. {\bf A55}, 1703 (1997).; A. Kent, gr-qc 9607073.

\bibitem{IL} C. J. Isham and N. Linden, quant-ph 9612035.

\bibitem{Pen} R.\ Penrose in { \sl Quantum concepts in space and
time}, edited by R.\ Penrose and C.\ J.\ Isham (Clarendon Press,
Oxford, 1986).


\bibitem  {ILS} C. J. Isham, N. Linden and S. Schreckenberg, J. Math. Phys. {\bf 35}, 6360 (1994).

\bibitem {GRW} G.\ C.\ Ghirardi, A.\ Rimini and T.\ Weber, {\sl
Phys. Rev. D}{\bf 34}, 470 (1986).


\bibitem {Per} Perelomov, {\sl Generalized coherent states and their applications}, (Springer-Verlag, Berlin- Heidelberg, 1986).

\bibitem{Ana} C. Anastopoulos, Phys. Rev. {\bf D 56}, 1009 (1997).

\bibitem{Mar} Margenau and Park, Int. Jour. Theor. Phys. {\bf 1}, 212 (1968).

\bibitem{Aus} M. Adelman and J. V. Corrett, {\sl A comparison of the closed subspace logic of Birkhoff-von Neumann with open subspace logic.}

\bibitem{Sor} R. D. Sorkin, Mod. Phys. Lett. {\bf A9}, 3119 (1994) ;
in {\sl Proceedings of the 4th Drexel Symposium on Quantum Nonintegrability:
Quantum Classical Correspondence}, edited by D.H. Feng and B. L. Hu , (International Press, 1996).



\end{thebibliography}
     \end{document}